\documentclass[%
reprint,
superscriptaddress,
 amsmath,amssymb,
 aps, pra
]{revtex4-1}

\usepackage[colorlinks=true,urlcolor=blue,citecolor=blue,linkcolor=blue]{hyperref}
\usepackage[T1]{fontenc}
\usepackage[latin9]{inputenc}
\usepackage{graphicx}
\usepackage{color}
\usepackage{mathrsfs}
\usepackage{indentfirst}
\usepackage{subfigure}
\usepackage{multirow}
\usepackage{booktabs}
\newcommand{\ra}[1]{\renewcommand{\arraystretch}{#1}}

\usepackage{placeins}
\usepackage{float}
\begin{document}

\title{Heavy tails in the distribution of time-to-solution for classical and quantum annealing}

\author{Damian S. Steiger}
\affiliation{Theoretische Physik, ETH Zurich, 8093 Zurich, Switzerland}
\author{Troels F. R{\o}nnow}%
\affiliation{Theoretische Physik, ETH Zurich, 8093 Zurich, Switzerland}
\affiliation{Nokia Technologies, Broers Building, 21 JJ Thomson Avenue, CB3 0FA
  Cambridge, United Kingdom}
\author{Matthias Troyer}
\affiliation{Theoretische Physik, ETH Zurich, 8093 Zurich, Switzerland}

\date{\today}% It is always \today, today,

\begin{abstract}
For many optimization algorithms the time-to-solution depends not only on the problem size but also on the specific problem instance and may vary by many orders of magnitude.  It is then necessary to investigate the full distribution and especially its tail. Here we analyze the distributions of annealing times for simulated annealing and simulated quantum annealing (by path integral quantum Monte Carlo) for random Ising spin glass instances. We find power-law distributions with very heavy tails, corresponding to extremely hard instances, but far broader distributions -- and thus worse performance for hard instances -- for simulated quantum annealing than for simulated annealing. Fast, non-adiabatic,  annealing schedules can improve the performance of simulated quantum annealing for very hard instances by many orders of magnitude.
\end{abstract}

\maketitle

%\tableofcontents
Non-convex optimization problems arise in a wide range of areas, from industrial applications to scientific research. Many challenging optimization problems belong to the class of non-deterministic polynomial-time hard (NP-hard) problems. For these problems, which includes the famous traveling salesman problem, no efficient algorithm is known that scales polynomially with problem size. In the absence of efficient exact solvers, often heuristic algorithms are the best choice. For many algorithms, and especially probabilistic annealing strategies \cite{Kirkpatrick1983,Suman2006,Kadowaki1998,Santoro2002,Santoro2006}, the time to find a (near)-optimal solution may depend not just on the problem size $N$, but may vary greatly -- by many orders of magnitude -- from instance to instance. In such cases one usually quotes the {\em typical} performance given by the median time-to-solution.

Despite its common use, the median time-to-solution is often not the quantity that dictates efficiency, since one generally wants to solve more than half of the problem instances. In particular, in the case of broad distributions of the time-to-solutions, higher quantiles (such as the 99-th percentile) may be better indicators of the performance of an optimization algorithm. In this Letter we demonstrate that when analyzing the performance of optimization algorithms, it is very important to
consider the tails of the time-to-solution distribution and not only averages or median performances.

 \begin{figure}[t]
 \centering
\includegraphics[width=\linewidth]{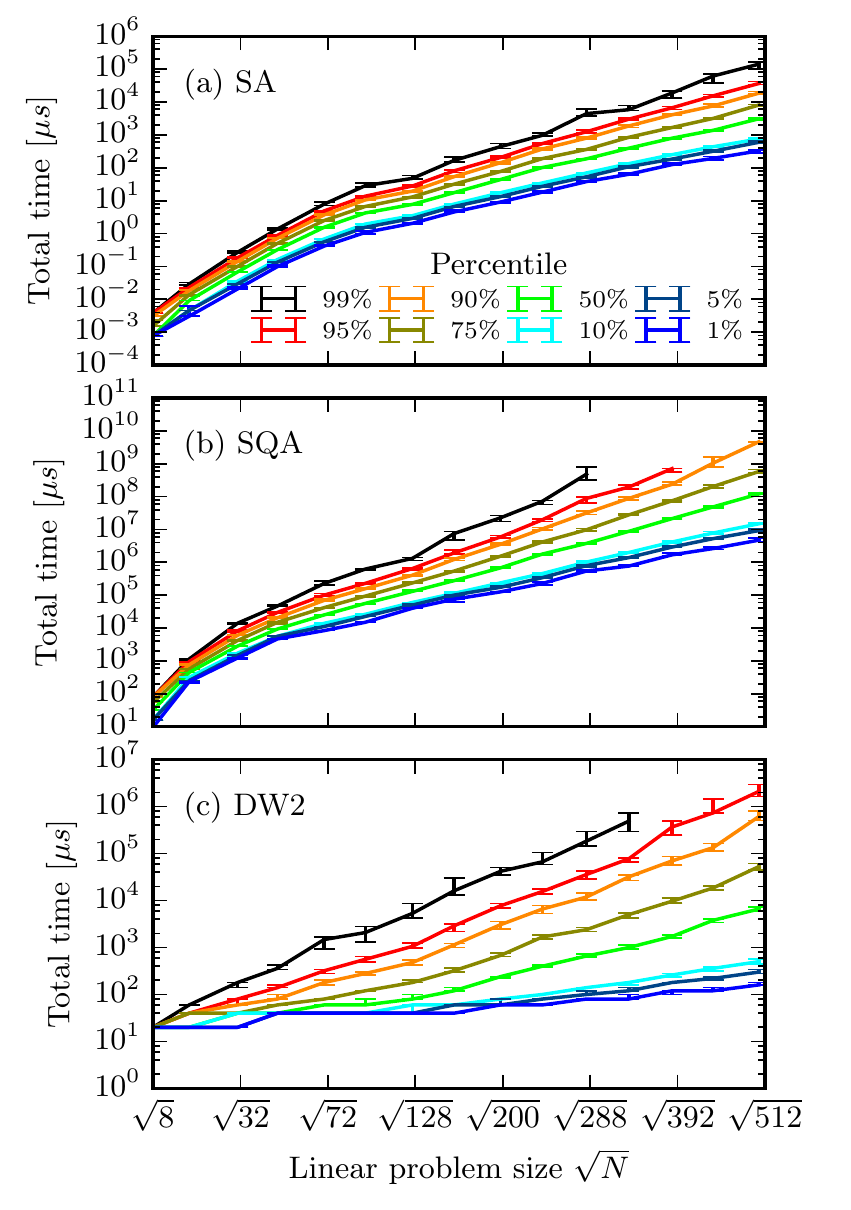}
 \caption{Scaling of the annealing time to find the ground state with 99\% probability for a bimodal Ising spin glass with couplings $\pm1$ on chimera graphs with $N$ spins using (a) a simulated annealer (SA), (b) a simulated quantum annealer (SQA) and (c) a D-Wave Two device (DW2). Results (with standard errors) are taken from Ref.~\cite{Ronnow2014}. We observe an enormous spread in the scaling of different percentiles, ranging from easy (1\%) to hard (99\%) problem instances. }
 \label{fig:previous_benchmark}
 \end{figure}

As a specific example we consider the problem of finding the ground states  of Ising spin glasses on chimera graphs. In these models $N$ Ising spins $s_i$ that can take values $\pm 1$ should be chosen to minimize the total energy,
\begin{equation}
\label{eq:IsingHamiltonian}
  H_P = -\sum_{\langle i,j\rangle} J_{ij} s_i s_j ,
\end{equation}
where  the coupling constants  $J_{ij}$ take random values on the edges of the so-called Chimera graph. For more information see the Supplementary Material. This quasi-two-dimensional graph with eight spins per unit cell is non-planar, which makes the problem of finding the ground state of the Ising spin glass  NP-hard \cite{Barahona1982}.

The Ising spin glass problem on the chimera graph has recently received special attention, since this graph is implemented in the optimization devices built by the Canadian company D-Wave systems  \cite{Johnson2011}, whose performance has been controversially discussed in the recent literature \cite{Boixo2014,Ronnow2014,Shin2014,McGeoch,Dash2013,Katzgraber:2014cy}. We summarize, in Fig.~\ref{fig:previous_benchmark} previous results \cite{Ronnow2014} of the scaling of the time-to-solution (TTS) as a function of problem size $N$ for three different approaches: a simulated annealer (SA), a simulated quantum annealer (SQA) using path integral quantum Monte Carlo, and a D-Wave Two device (DW2). 
Note one striking feature of these results: there is a huge spread in TTS between easy and hard instances (the 99th percentile), which varies by three orders of magnitude for SA, and by even more for SQA and DW2. The hard instances completely dominate the TTS. Below we will quantitatively analyze the distribution of TTS for the hardest instances, and show how it can be substantially reduced for SQA by using non-adiabatic annealing schedules.

Simulated annealing (SA) is inspired by annealing, a process in metallurgy by which a material is 
heated up and then slowly cooled down in order to relieve internal stresses.
In 1983, Kirkpatrick and coworkers \cite{Kirkpatrick1983} showed that simulating an annealing process a Monte Carlo simulation can be used as a general purpose heuristic solver for optimization problems, which is widely used in many application areas \cite{Suman2006}. By slowly decreasing the temperature the system can relax into a low-energy state, escaping local minima by thermal activation over barriers. Implementation details are discussed in the Supplementary Material.

Inspired by the phenomenon of quantum tunneling, quantum annealing (QA)  uses quantum 
fluctuations instead of thermal fluctuations to explore the configuration space \cite{Ray1989,Finnila1994,Kadowaki1998,Farhi2001,qareview}. It employs
a  time-dependent Hamiltonian,
\begin{equation}
  \label{eq:quantum_annealing}
  H(t) = A(t)H_D + B(t)H_P,
\end{equation}
where $H_D = -\sum_i\sigma_i^x$
 is a driver Hamiltonian implementing quantum dynamics, and the problem Hamiltonian
$  H_P = -\sum_{\langle i,j\rangle} J_{ij} \sigma_i^z\sigma_j^z$ implements the Ising spin glass Eq.~\eqref{eq:IsingHamiltonian} using quantum spin-1/2 particles. $\sigma^x$ and $\sigma^z$ are Pauli matrices. At the beginning of the annealing schedule when $A(0) = 1$ and $B(0)=0$ we start with only $H_D$. The spins are initialized in the ground state aligning along the x-axis. During the annealing process we decrease $A(t)$ and increase $B(t)$, so that at the end of the annealing at time $t=t_a$, we end up with the Ising spin glass of Eq.~\eqref{eq:IsingHamiltonian}. The Hamiltonian is changed slowly enough such that the system stays in (or close to) the ground state at all times. This process takes place at a constant finite temperature $T$ and therefore constant inverse temperature $\beta=1/(k_B T)$, where $k_B$ is the Boltzmann constant.
The D-Wave devices have been designed to implement quantum annealing using superconducting flux qubits  \cite{Johnson2011} in a physical device. 

Quantum annealing can also be implemented as simulated quantum annealing (SQA) on a classical computer using a path integral quantum Monte Carlo simulation \cite{Santoro2002,Santoro2006,Heim2015}. Furthermore, by replacing the quantum spins  in Eq.~\eqref{eq:quantum_annealing} with two-dimensional classical rotor magnets a mean-field version (MFA) \cite{Shin2014} can be obtained as a semi-classical version of  SQA \cite{Tameem2014}. Details of the SQA and MFA algorithms and their implementations are presented in the Supplementary Material. Note that annealing times $t_a$ for SA, SQA and MFA are measured in numbers of sweeps, where one sweep is defined as one attempted update   
for each spin.

\begin{figure}
  \centering
  \includegraphics[width=\linewidth]{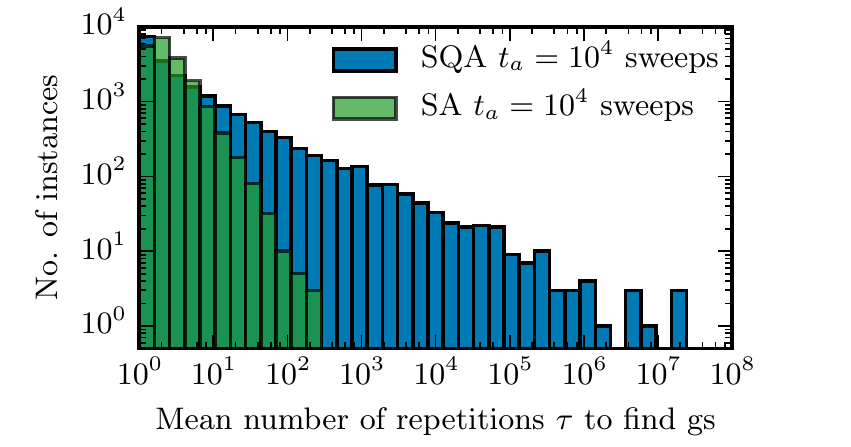}
  \caption{Distribution of mean number of repetitions $\tau$ required to find the ground state for 20000
    problems with 200 spins on Chimera topology for SA and SQA(with $\beta=10$). The tails of these distribution
    functions are decaying polynomially slow and there is a qualitative difference between SA and SQA:
    For SA the tail is decaying faster and the overall distribution function is more compact.} 
  \label{fig:distrib_SA_SQA}
\end{figure}

For each problem size $N$ we created 20000 different random problem instances with couplings $J_{ij}=\pm1$. For each instance, after determining the exact ground state energy using an exhaustive search algorithm we repeatedly performed SA, SQA and MFA simulations until we found a ground state at least 100 times for each algorithm. For SQA with $N=288$ spins and annealing time $t_a=10^4$ sweeps we never found the ground state of the hardest instances despite billions of attempts. We thus focus on the case of  $N=200$ spins for our quantitative analysis, where all algorithms found the ground states of all problem instances with reasonable effort. 
From the number of repetitions required to find the ground state one hundred times we then calculated the single-run success probability $s$ and the mean number of repetitions $\tau=1/s$ required to find the ground state. 

Plotting histograms of $\tau$ for 20000 instances with $N=200$ we arrive at the distributions shown in Fig.~\ref{fig:distrib_SA_SQA}. We see very slowly decaying power-law tails, consistent with the wide spread of quantiles in Fig.~\ref{fig:previous_benchmark}. We also see that while the median times are very similar ($\tau=2.1$ for SA and $\tau=2.2$ for SQA), there is a substantial quantitative difference between the distributions. For SA tails extend up to large values of $\tau\approx100$, but for SQA to values of $\tau\approx10^7$ which is more than 5 orders of magnitude larger. The median times are thus neither indicative of the time needed to solve hard problem instances, nor of the relative performance of 
the two algorithms!

To analyze this significant difference in the tail distribution function in detail, we use 
extreme value theory (EVT) to obtain a good estimate for the tail distribution 
of $\tau$ given only a limited data set and specifically to calculate the so-called shape parameter $\xi$, which gives information about its power-law tail.
Let us denote by 
$F(x)=\mathrm{Pr}\left\lbrace X \leq x \right\rbrace$ the unknown distribution function of a random variable $X$, e.g. the distribution of $\tau$ for a specific algorithm.
If $F$ satisfies certain conditions (described in the Supplementary Material)
then the Balkema-de Haan-Pickands (BdHP) theorem \cite{Balkema1974,Pickands1975}
states that the tail of the distribution 
function, $F(x)$ for $x$ larger than a high threshold $u$, is approximated by a generalized Pareto (GP) distribution function $W_{\xi , \tilde{u}_u , \tilde{\sigma}_u}(x)$, i.e.
\begin{equation}
F(x)\approx W_{\xi , \tilde{u}_u , \tilde{\sigma}_u}(x)\quad \mathrm{for}\; x\geq u.
\end{equation}
The GP distribution function with threshold/location parameter $u$, shape parameter $\xi$, scale parameter
$\sigma_u>0$ is given by
\begin{equation}
W_{\xi,u,\sigma_u}(x):=1-\left(1+\xi \left(\frac{x-u}{\sigma_u}\right)\right)^{-1/\xi} ,
\end{equation}
which is defined on $\left\lbrace x : \, x \geq u \; \mathrm{and} \; 
\left( 1+\xi (x-u) / \sigma_u \right)\geq 0\right\rbrace$ \cite{Reiss2007,Falk2011,Coles2001}.
For $\xi=0$ one takes the limit $\xi\to0$,
\begin{equation}
W_{0,u,\sigma_u}(x)=1-\exp\left(-(x-u)/ \sigma_u\right),\quad x \geq u .
\end{equation}

 \begin{figure}
 \centering
 \includegraphics[width=\linewidth]{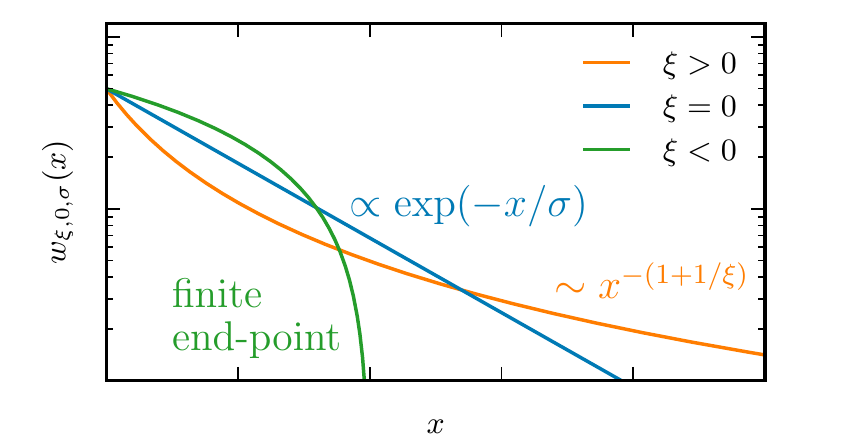}
 \caption{Generalized Pareto probability density function $w_{\xi,0,\sigma}(x)=\frac{d}{dx}W_{\xi,0,\sigma}(x)$ plotted on a linear-log scale. The qualitative 
 tail behavior  depends on the shape parameter $\xi$. For $\xi<0$ 
 the probability density is bounded, for $\xi=0$ it decays
 exponentially and for $\xi>0$ the tail is heavy and decays as a power-law.}
 \label{fig:gpd_density}
 \end{figure}
 
The qualitative behavior of the GP distribution function is given by its shape 
parameter $\xi$, which turns out to be very useful when benchmarking 
different algorithms. The shape parameter is independent of the 
threshold $u$ and of any linear transformations on $X$, e.g.
considering the number of spin flips to find the ground state instead of $\tau$. 
The  probability density has three distinct tail behaviors, shown in
Fig.~\ref{fig:gpd_density}, depending on the sign of $\xi$.
It is bounded for $\xi<0$, exponentially decaying for $\xi=0$ and slowly decaying as a power-law for $\xi>0$, which we also refer to as having a heavy tail.
Note that the $k$-th moment of 
$W_{\xi,u,\sigma_u}(x)$ diverges for $k\geq 1/\xi$ \cite{McNeil2005}.

If a heuristic algorithm has $\xi>0$, 
then hard instances dominate the average time-to-solution. If
the shape parameter $\xi\geq1$ the average run-time of an instance is infinite.
In our case of discrete couplings $\pm1$, there is only a  finite
set of instances for each $N$. The conditions of the the BdHP theorem are thus not satisfied because the distribution  of $\tau$ is discrete with only a finite number of points in its support \cite[p. 217]{Arnold2008}. Nevertheless we can still use the GP distribution to analyze our data, and obtain excellent fits (see Supplementary Material). The main consequence of a finite set of instances is that $F(\tau)$ differs from the GP distribution function
for the hardest instances, and thus for example the mean $\tau$ is not infinite but dominated by the hardest instances. In the Supplementary Material we show that the mean is not converging for $\xi\geq1$ using 20000 problem instances.

\begin{figure}[t]
  \centering
   \includegraphics[width=\linewidth]{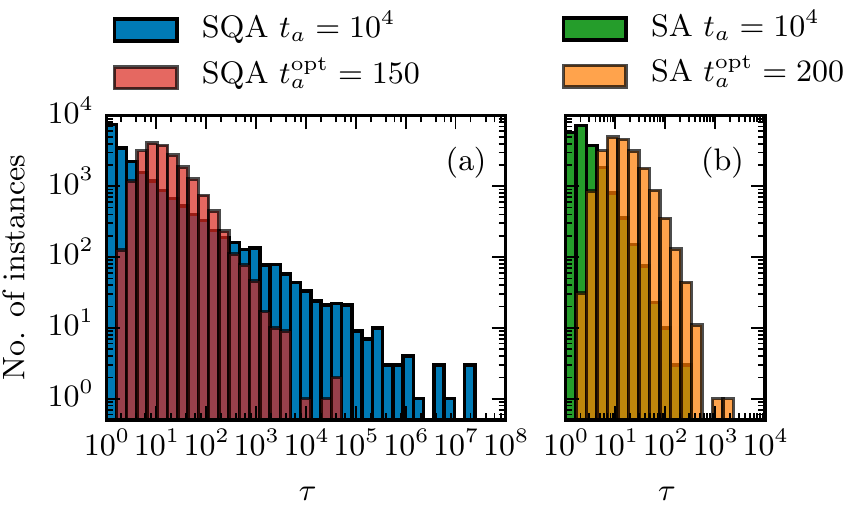} 
  \caption{Distribution of mean number of repetitions $\tau$ required to find the ground state for 20000
    problem instances with $N=200$ spins for (a) SQA with $\beta=10$ and (b) SA with 
    either the optimal number or $10^4$ sweeps.   While the distribution changes only slightly for SA, surprisingly the tails in SQA decrease much more rapidly when annealing faster. } 
  \label{fig:distrib_SA_SQA_comparing}
\end{figure}

\begin{figure}
  \centering
  \includegraphics[width=\linewidth]{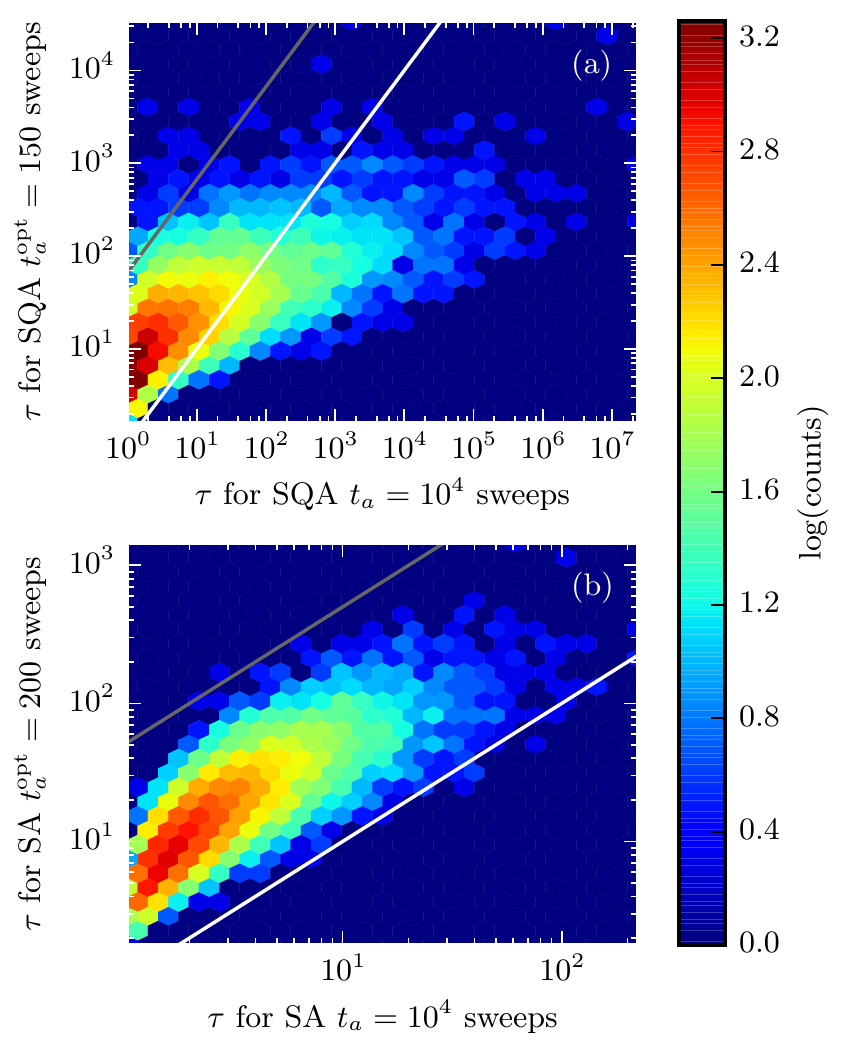} 
  \caption{Correlation plot of (a) SQA with $\beta=10$ and (b) SA for all 20000 instances 
  with $N=200$ spins using $t_a^{\rm opt}$ and $t_a=10^4$ sweeps respectively.  Points below the white line ($\tau$,$\tau$) are those where the success probability $s$ increases upon annealing faster. This is the case for 2312 instances for SQA but only 11 for SA. This increase in $s$ for the hard instances explains the more rapid decrease of the tail in SQA. The grey line indicates points with equal total effort $ t_a\tau$. Instances below the grey line (all but one for SQA and all but 51 for SA) are those which get solved more efficiently with a faster annealing time.} 
  \label{fig:SA_SQA_correlation}
\end{figure}

We start, by investigating the effects of annealing time $t_a$ on the distribution. In  Fig.~\ref{fig:distrib_SA_SQA_comparing} we  compare the distributions for $t_a=10^4$ sweeps (which for SQA with $\beta=10$ was found to give good correlations with a D-Wave One device \cite{Boixo2014}) to simulations run with the (size-dependent) optimal $t_a^{\rm opt}$ that minimizes the total effort $t_a\tau$ (see Ref. \onlinecite{Ronnow2014} and the Supplementary Material). The surprising result is that performing SQA with a fast schedule of only $t_a^{\rm opt}=150$ sweeps leads to a more compact distribution with much faster decaying tails. While easy instances may need a few more repetitions, hard instances need far less repetitions to find the ground state. 

As the correlation plot in Fig.~\ref{fig:SA_SQA_correlation} shows, the relative hardness of an instance doesn't change significantly when changing $t_a$, but the single-run success probability $s$ of SQA increases for more than 10\% of the instances when SQA is run faster. This is opposite to SA where for almost all instances $s$ decreases when annealing faster. In both cases, however, the total effort $\tau t_a$ decreases for almost all instances. Even though more repetitions $\tau$ are needed, this is more than compensated by the faster annealing time. For SQA, the combined effect of needing 688 times fewer repetitions and performing only 150 instead of $10^4$ sweeps per repetition reduces the total effort for the hardest of our instances by a factor 45866!

\begin{figure}
  \centering
  \includegraphics[width=\linewidth]{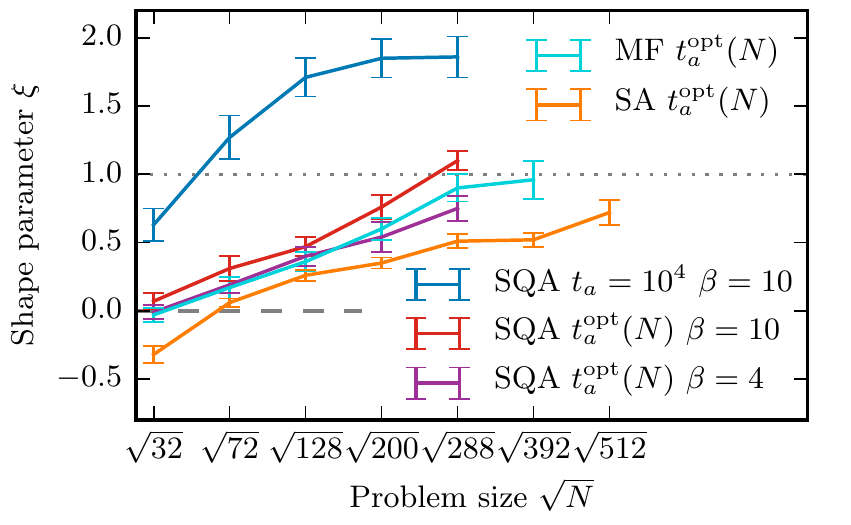} 
  \caption{Shown are the fitted shape parameters $\xi$ with standard errors for SA, MFA and SQA. The different versions of SQA have different inverse
  temperature $\beta$ and $t_a$, with $t_a^{\rm opt}(N)$ varying from 20 to 400
   sweeps depending on $N$ and $\beta$. For $\xi>0$ (indicated by a dashed line) the tail of the 
   distribution function is heavy and for $\xi\ge1$ (indicated by a dotted line) the mean diverges. The $k$th moment diverges for $k\geq1/ \xi$.} 
  \label{fig:shape_param}
\end{figure}

To quantify the difference in the tail distribution functions we fit them to GP distribution functions and plot the size-dependence of the shape parameter $\xi$ for SA, SQA and MFA in Fig.~\ref{fig:shape_param}.  For SQA we use the two schedules discussed above and additionally a third schedule, with $t_a^{\rm opt}(N)$ sweeps but a higher temperature $\beta=4$, which is the optimal temperature for  MFA (see Supplementary Material).
All algorithms have heavy tails, $\xi>0$, for problems with size $N\geq72$, and $\xi$ is increasing with system size. For SQA using  $\beta=10$ and $t_a=10^4$ we find $\xi\geq1$ already for $N\ge72$, indicating that the mean time-to-solution is already divergent and dominated by the hardest instances. As we already saw above for $N=200$, the tail behavior of SQA improves when using faster annealing schedules, and further small improvements are obtained by raising the temperature. While SQA with the best setting and MFA have the same $\xi$, SA has  smaller $\xi$, indicating a faster decaying tail and better performance on hard instances.

Our results show that the tail behavior is the important measure of performance when 
considering the performance of an optimization algorithm since the hardness of an instance is a priori unknown. Therefore one has to choose a large
enough run-time such that a high percentile of random instances are solved.
Considering just the typical (median) performance of an optimization algorithm can be misleading in the presence of broad heavy-tailed distributions. The median effort misses the important fact that already at moderate problem sizes the mean time-to-solution can diverge, indicating that the hardest instances dominate the average. As one typically wants to solve much more than half of the problem instances, reducing the tails of the distribution is crucial to optimize the performance of an optimization algorithm. 

One surprising result we obtained here is that for simulated quantum annealing, fast non-adiabatic schedules show far superior performance than slow adiabatic annealing schedules that would na\"ively seem superior. This observation is consistent with results obtained in Refs. \cite{Heim2015,Crosson:2014ez} which  showed that for some instances fast annealing schedules can help a (simulated) quantum annealer escape local minima. As good agreement was found between SQA and the D-Wave devices, we expect that similarly faster annealing schedules can help the performance not only of simulated quantum annealers but also of physical quantum annealing devices. 

The tail behavior is also important for extrapolating the performance of annealers to larger sizes. The tails are indicative of harder problems that will dominate for larger problem sizes. We thus expect scaling differences between different annealers to first show up in the tails.

%\textit{Acknowledgements.}
We acknowledge discussions with Sergio Boixo, Paul Embrechts, Hartmut Neven and Diethelm W\"urtz and thank Ethan Brown for proofreading and Coal Ila for inspiration. This work has been supported by the Swiss National Science Foundation through the National Competence Center in Research QSIT and is based upon work supported in part by 
%the Office of the Director of National Intelligence (ODNI), 
%Intelligence Advanced Research Projects Activity (IARPA)
%ODNI, 
IARPA
via MIT Lincoln Laboratory Air Force Contract No. FA8721-05-C-0002. 
%The views and conclusions contained herein are those of the authors and should not be interpreted as necessarily representing the official policies or endorsements, either expressed or implied, of ODNI, IARPA, or the U.S. Government.  
% The U.S. Government is authorized to reproduce and distribute reprints for Governmental purpose notwithstanding any copyright annotation thereon.

\bibliography{ref_paper}% Produces the bibliography via BibTeX.

%%%%%%%%%% Merge with supplemental materials %%%%%%%%%%
\pagebreak
\pagebreak
%\widetext
\onecolumngrid
\begin{center}
\textbf{\large Supplementary Material for ``Heavy tails in the distribution of time-to-solution for classical and quantum annealing''}
\end{center}
\hspace{0.5cm}
\twocolumngrid
%%%%%%%%%% Merge with supplemental materials %%%%%%%%%%
%%%%%%%%%% Prefix a "S" to all equations, figures, tables and reset the counter %%%%%%%%%%
%\setcounter{equation}{0}
%\setcounter{figure}{0}
%\setcounter{table}{0}
%\setcounter{page}{1}
%\makeatletter
%\renewcommand{\theequation}{S\arabic{equation}}
%\renewcommand{\thefigure}{S\arabic{figure}}
%\renewcommand{\bibnumfmt}[1]{[S#1]}
%\renewcommand{\citenumfont}[1]{S#1}
%%%%%%%%%% Prefix a "S" to all equations, figures, tables and reset the counter %%%%%%%%%%

\section{Annealing algorithms}
In the following we present implementation details for the three annealing algorithms.

\subsection{Simulated annealing}

SA is a Monte Carlo algorithm which generates states of a classical system in thermal equilibrium whose energy is described by the following Hamiltonian
\begin{equation}
\label{eq:IsingHamiltonian_SM}
  H_P = -\sum_{\langle i,j\rangle} J_{ij} s_i s_j .
\end{equation}
Our SA code generates new states by flipping a single spin and accepting the spin flip according to the Metropolis update rule \cite{Metropolis1953}. Hence, the probability to generate a state with energy $E$ is proportional to the Boltzmann weight $\exp(-\beta E)$, where $\beta=1/(k_B T)$ is the inverse temperature. Over the annealing time $t_a$, which is measured in numbers of sweeps (one sweep is an attempted flip of every spin in the system), the temperature is slowly decreased. In the beginning, when the temperature is high, SA has a large probability to generate high energy states, called thermal excitations, which allow SA to escape from a local minimum. In the end,
the temperature is low such that mostly states with lower energy are accepted and the system relaxes into the local minimum. For these instances with $J_{ij}=\pm1$, we linearly increase the inverse temperature $\beta(t)=0.1+2.9\times t/t_a$ over the annealing time $t_a$. In order to reduce CPU time, our implementation uses forward computation of the energies as described in Ref. \cite{Isakov2014}.

\subsection{Simulated quantum annealing}

SQA is a Monte Carlo algorithm which generates states of a quantum system in thermal equilibrium. During the annealing the temperature is kept constant. The Hamiltonian of the system is time dependent and given by Eq.~\eqref{eq:quantum_annealing} in the main paper. At the beginning of the annealing schedule when $A(0) = 1$ and $B(0)=0$ we start with only $H_D$. We use a linear schedule, i.e. we linearly decrease $A(t)=1-t/t_a$ and increase $B(t)=t/t_a$, so that at the end of the annealing at time $t=t_a$, the Hamiltonian of the system is given by $H_P$.
SQA uses the Suzuki-Trotter formalism with which a $d$-dimensional Ising model with a transverse field is mapped
to a ($d+1$)-dimensional Ising model at finite temperature \cite{Suzuki1976}. Here we use a discrete-time path integral formulation with 32 imaginary time slices and we
build clusters along the imaginary time direction. On a technical level, we exploit a random bit algorithm explained in Ref. \cite{Pierre1987}
to form new domain walls in all 32 Trotter slices simultaneously.\\

We analyze SQA as simulation of a physical system, which means that in the end of annealing, the success probability of finding the ground state is given by the number of Trotter slices which reached the ground state divided by 32 (the total number of Trotter slices). This means for example that if half the Trotter slices are in the ground state, we count this run as having found the ground state only with probability 0.5, because an actual quantum system would collapse into one of the Trotter slices. One could on contrary also analyze SQA as a purely classical algorithm and define that a single run has found the ground state energy with probability 1 if at least one Trotter slices is in the ground state, as one can check all Trotter slices in a classical code. This way of analyzing SQA resulted in very similar shape parameters $\xi$ of the fitted GP distribution functions for $\tau$, which means there is no qualitative difference in the tail behavior and therefore we only show the results of analyzing SQA as a simulation of a physical system throughout the whole paper.

\subsection{Mean-field quantum annealing}
MFA is a mean-field version of simulated quantum annealing. A spin $i$ is represented by a classical magnet pointing in some direction $\theta_i$ in the XZ plane \cite{Shin2014}. The classical time-dependent Hamiltonian is then given by:
\begin{equation}
 H(t) = -A(t)\sum_i \sin\left(\theta_i\right) - B(t)\sum_{\langle i,j\rangle} J_{ij} \cos\left(\theta_i\right)\cos\left(\theta_j\right)
\end{equation}
MFA is a Monte Carlo algorithm which generates states of this system at thermal equilibrium and the temperature is again kept constant.
A new state is generated by proposing an update of spin $i$ by choosing uniform randomly a new angle $\theta_i\in [0,2\pi)$ and then accepting or rejecting this update according to the Metropolis update rule \cite{Metropolis1953}.
We use the same linear schedule for the constants $A(t)$ and $B(t)$ as in the case for SQA and all spins are initialized to be aligned in x-direction at $t=0$. Our MFA implementation follows the description in Ref. \cite{Shin2014}, but differs by using 1024 element lookup tables with precomputed values of
$\sin(\theta)$ and $\cos(\theta)$.

\section{Problem instances}

\begin{figure}[t]
  \centering
  \includegraphics[width=0.3\textwidth]{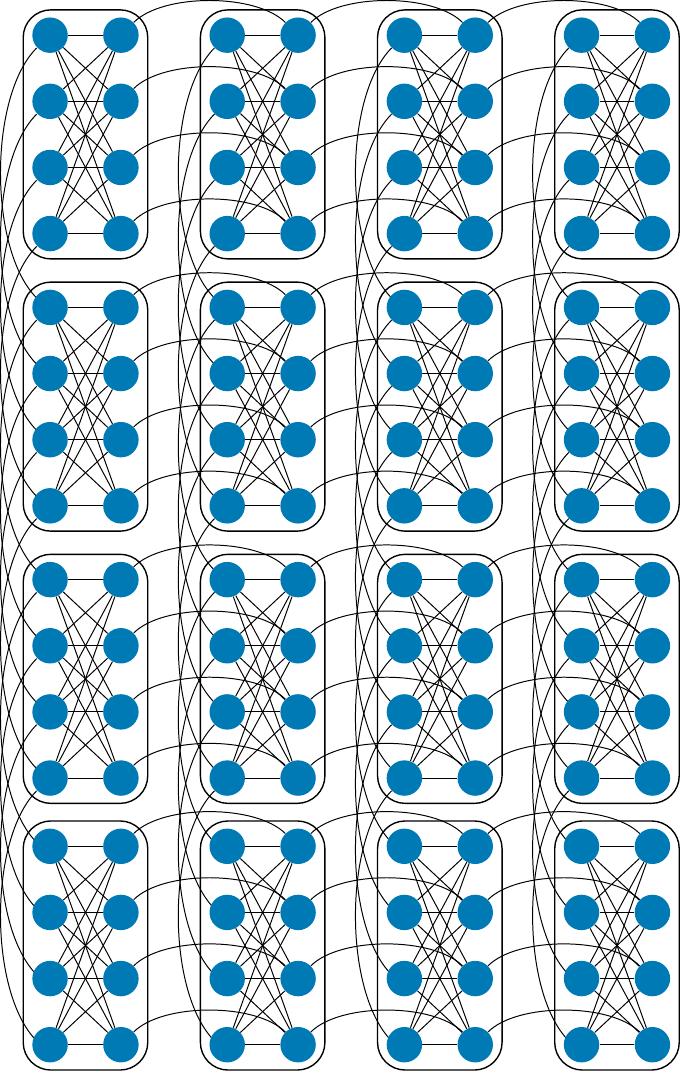}
  \caption{{\bf Chimera graph.} Shown is a 128 spin Chimera graph
    consisting of $4\times4$ unit cells each containing 8 spins.}
  \label{fig:figure_chimera_graph}
\end{figure}

The present analysis was carried out on Ising spin glasses with Chimera
topology. The Chimera graph is formed by a two-dimensional lattice built from eight-spin unit cells that each form a complete bipartite graph $K_{4,4}$. Fig.~\ref{fig:figure_chimera_graph} shows the chimera graph for 128 spins, corresponding to $4\times 4$ unit cells.
We used graphs with $L \times L$ unit cells and $8L^2$ spins, with $L$ ranging from 2 to 8.
The couplings between the spins were randomly chosen to be $\pm1$.

To find the ground states of these Ising spin glasses, we implemented an exact solver using dynamic programming. The solver exploits the limited tree-width of the Chimera graph which allows one to perform an exhaustive search up to 512 spins.

Note that since Chimera graphs are bipartite, all our Monte Carlo annealing algorithms
could potentially be parallelized efficiently because spins in one sublattice couple only to spins in the other sublattice and hence all spins in one sublattice can be updated simultaneously.

\section{Optimal number of sweeps}

\begin{figure}[t]
  \centering
  \includegraphics[width=\linewidth]{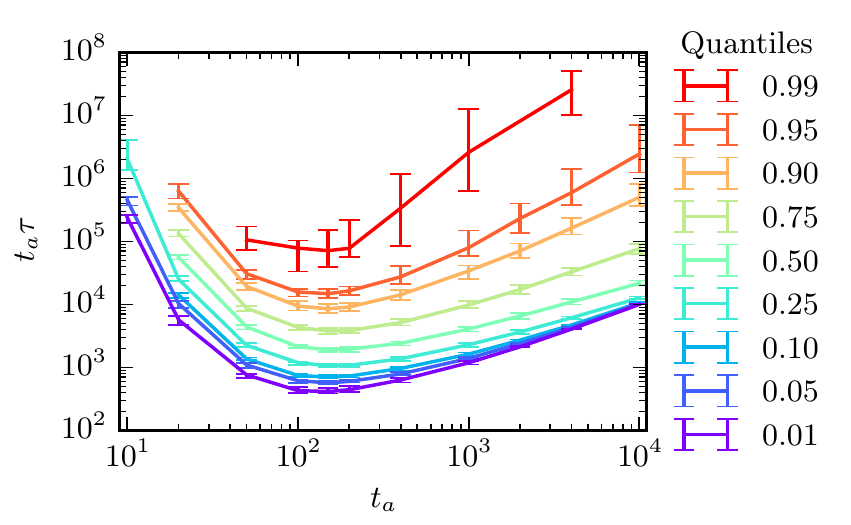} 
  \caption{Choosing the optimal annealing time per run $t_a^{\mathrm{opt}}(N)$ for SQA at $\beta=10$ and problems with $N=200$ spins. 
  The single-run success probability $s$ is estimated for 1000 random instances for SQA with different annealing times $t_a$ by making 12800 repetitions for each instance and $t_a$. Then the mean number of repetitions $\tau=1/s$ required to find the ground state is calculated and the total effort $t_a\tau$ to find the ground state is plotted (with $95\%$ confidence intervals). The optimal annealing time $t_a^{\mathrm{opt}}$ is the $t_a$ which minimizes $t_a\tau$ for most quantiles. For this case $t_a^{\mathrm{opt}}=150$ sweeps. Note that the minimum of each quantile is at the same $t_a$ and very flat.} 
  \label{fig:opt_sweeps}
\end{figure}

When investigating the scaling of annealers with problem size $N$ one has to carefully choose optimal parameters or one might draw wrong conclusions \cite{Boixo2014,Ronnow2014}. One could,
for instance, choose to run the annealer $N$ times and each time anneal
for time $t_a = t$ and then choose the lowest lying state found, or, 
as an extreme case example, one 
could choose to make a single annealing run over a time span $t_a=N\cdot t$. The
result in the two cases are very different, even though the total annealing time
is equal in both cases.

To make extrapolations to asymptotic scaling one needs to use an optimal annealing time $t_a^{\rm opt}$, which
is defined to be the number of sweeps $t_a$ which minimizes
the total effort $t_a\tau$, where $\tau$ is the mean number of repetitions required to find the ground state. 
See Fig. \ref{fig:opt_sweeps}. 
Note that the minima for the different quantiles are usually at roughly the same $t_a$ and the minima are  quite broad and flat. 
Thus $t_a^{\rm opt}$ does not have to be determined with very high accuracy.
The optimal annealing times $t_a^{\rm opt}(N)$ are listed in Tab.~\ref{tab:opt_num_sweeps} for SA, SQA and MFA.

\begin{table} %optimal sweeps MF
\ra{1.3}
\begin{center}
\begin{tabular}{@{}rcrcrrcrr@{}}\toprule
 &\phantom{ab}& \multicolumn{1}{c}{$t_a^{opt}$ of SA}&\phantom{ab}& \multicolumn{2}{c}{$t_a^{opt}$  of SQA}&\phantom{ab}&\multicolumn{2}{c}{$t_a^{opt}$  of MFA}\\ \cmidrule{3-3} \cmidrule{5-6} \cmidrule{8-9}
$N$&&&&$\beta=10$&\phantom{b}$\beta=4$&&$\beta=10$&\phantom{a}$\beta=4$\\ \midrule
32 && 10 && 20 & 20 && 20 & 50\\
72 && 25 && 50 & 50 && 100 & 100\\
128&& 70&&100&100&&500&200\\
200&&200&&150&150&&2000&1000\\
288&&300&&200&400&&5000&2000\\
392&& 650&&400&1000&&   &5000\\
512&&1500&&400&1000&&   &5000\\
\bottomrule
\end{tabular}
\end{center}
\caption{ Optimal annealing time $t_a^{\mathrm{opt}}$ (given in numbers of sweeps) for SA, SQA or MFA to solve Ising spin glass problems on Chimera graphs with $\pm1$ couplings and $N$ spins. For SQA and MFA we estimated the optimal annealing time for different inverse temperatures $\beta$.}
\label{tab:opt_num_sweeps}
\end{table}

\section{Optimal temperature of MFA}
\begin{figure}[t]
  \centering
  \includegraphics[width=\linewidth]{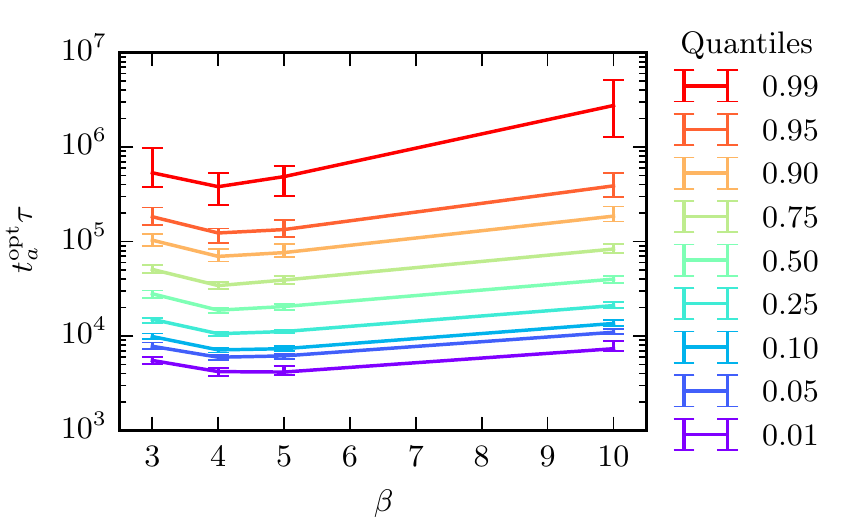} 
  \caption{Determining the optimal inverse temperature $\beta$ for MFA. Plotted are the quantiles of the total effort, $t_a^{\mathrm{opt}}\tau$, required to find the ground state (with $95\%$ confidence intervals) for different values of $\beta$. At each $\beta$ the optimal annealing time $t_a^{\mathrm{opt}}$ is calculated and used in this plot. The optimal inverse temperature is $\beta=4$.} 
  \label{fig:opt_temp_mfa}
\end{figure}

For the MFA algorithm we determined that the optimal inverse temperature is $\beta=4$ which minimizes the total effort required to find the ground state, $t_a\tau$, for most quantiles and problem sizes $N$. See Fig.~\ref{fig:opt_temp_mfa} which shows the case for $N=200$.

%\FloatBarrier
\section{Extreme Value Theory}
\label{sec:EVT}
In this section we summarize the important parts of Extreme Value Theory (EVT) which are used in the main paper. A detailed discussion can be found in the cited papers and textbooks.
Here we will state the Balkema-de Haan-Pickands (BdHP) theorem, which says that if $F(x)$ satisfies the extreme value condition, then $F(x)\approx W_{\xi , \tilde{u}_u , \tilde{\sigma}_u}(x)$ for $x\geq u$.
We will then discuss why our $F(x)$ doesn't satisfy the extreme value condition, but show that nevertheless our empirical data indicates that $F(x)\approx W_{\xi , \tilde{u}_u , \tilde{\sigma}_u}(x)$ for $x\geq u$ is a good model for the range of $x$ which we are interested.

\subsection{Theory}
Let $X$ be a random variable, i.e. the run-time of an instance using a specific algorithm,
 which is described by the unknown distribution function 
$F(x)=\mathrm{Pr}\left\lbrace X \leq x \right\rbrace$.

Then the exceedance distribution function over a threshold 
$u$, $F^{\left[ u \right]}$, is defined as
\begin{equation}
\label{eq:exceedance_dist}
F^{\left[ u \right]}(x):=\mathrm{Pr}\left\lbrace X \leq x \mid X>u \right\rbrace=
\frac{F(x)-F(u)}{1-F(u)}\; \mathrm{for}\; x\geq u.
\end{equation}
The Balkema-de Haan-Pickands theorem\cite{Balkema1974,Pickands1975}
now says that if $F^{\left[ u \right]}\left(b_u+a_u x\right)$ has, for some
choice of constants $b_u$ and $a_u$, a continuous limiting distribution 
function as $u$ goes to the right endpoint $\omega(F)$ of F, which is also called
the extreme value condition \cite{Reiss2007}, then:
\begin{equation}
\nonumber
\left| F^{\left[u\right]}(x)-W_{\xi,u,\sigma_u}(x)\right|\to 0, \;\;\mathrm{for}
\; u\to \omega(F),\; x\geq u.
\end{equation}
with threshold/location parameter $u$, shape parameter $\xi$, scale parameter
$\sigma_u>0$ and $W_{\xi,u,\sigma_u}(x)$ is a generalized Pareto (GP) distribution
function
\begin{equation}
W_{\xi,u,\sigma_u}(x):=1-\left(1+\xi \left(\frac{x-u}{\sigma_u}\right)\right)^{-1/\xi} ,
\end{equation}
which is defined on $\left\lbrace x : \, x \geq u \; \mathrm{and} \; 
\left( 1+\xi (x-u) / \sigma_u \right)\geq 0\right\rbrace$ \cite{Reiss2007,Falk2011,Coles2001}.
For $\xi=0$ one has to take the limit $\xi\to0$: 
\begin{equation}
W_{0,u,\sigma_u}(x)=1-\exp\left(-(x-u)/ \sigma_u\right),\quad x \geq u .
\end{equation}
For a large enough threshold $u$ this theorem provides a parametric model 
for the exceedance distribution function 
$F^{ \left[ u \right] }(x) \approx W_{\xi , u , \sigma_u }(x)$. The parameters $\xi$ and $\sigma_u$ can be estimated by fitting the exceedances to $W_{\xi , u , \sigma_u }$ as explained later. If $F^{ \left[ u \right] }(x) \approx W_{\xi , u , \sigma_u }(x)$ is a good approximation, then it follows from the definition of $F^{[u]}(x)$ that \cite{Reiss2007}
\begin{equation}
 F(x)\approx W_{\xi , \tilde{u}_u , \tilde{\sigma}_u}(x)\quad \mathrm{for} \; x\geq u \;\mathrm{with}
\end{equation} 
 \begin{equation}
 \tilde{\sigma}_u=\sigma_u / \left( 1 + \xi W_{\xi,0,1}^{-1}\left( F(u) \right) \right)\quad \mathrm{and}
 \end{equation}
 \begin{equation}
 \tilde{u}_u=u-\tilde{\sigma}_u W_{\xi,0,1}^{-1}\left( F(u) \right).
 \end{equation}
The shape parameters $\xi$ of the GP distribution function which approximates $F^{[u]}(x)$ or $F(x)$ are identical. And in contrast to the scale parameter $\sigma_u$ which is dependent on the chosen threshold $u$, the shape parameter is independent of $u$, i.e. suppose that the threshold $u$ is high enough such that $F^{ \left[ u \right] }(x) \approx W_{\xi , u , \sigma_u }(x)$ then if exceedances over a higher threshold $\mu>u$ are fitted, it results in the same shape parameter $\xi$ because of the 
peak-over-threshold stability of GP distribution functions :  
$W_{\xi,u,\sigma}^{[\mu]}=W_{\xi,\mu,\sigma+\xi(\mu-u)}$ with 
$\sigma+\xi(\mu-u)>0$ \cite{Reiss2007}. 

In practice GP distribution functions are used as the canonical distribution function for modeling exceedances over a high threshold, see for example \cite{McNeil2005}.
Moreover, as the distribution function of interest $F(x)$ is unknown, it is commonly assumed as a null-hypothesis that $F(x)$ satisfies the extreme value condition and hence it can be approximated by a GP distribution function above a high threshold. Tests on the data and fitted model should be performed to potentially reject the null-hypothesis, which we introduce in the next subsection.
The estimated GP distribution functions are often used to make predictions of rare events, which are even more extreme than what has already been observed and used for the fitting of the GP distribution parameters. For example, extreme value analysis is used to plan the heights of sea dikes in the Netherlands such they fail no more than once in 10'000 years on average, but the data for the model fitting consists of the sea level from the last 100 years only \cite{Haan1990}.
In our case the extreme value condition is not satisfied, because as already stated in the main paper, our distribution functions of $\tau$ are discrete with only a finite number of points in its support, and therefore the extreme value condition doesn't hold, see \cite[p.217]{Arnold2008}. Nevertheless we assume as a null-hypothesis that $F(x)\approx W_{\xi , \tilde{u}_u , \tilde{\sigma}_u}(x)$ in the range of $x$ above the threshold $u$ which we are interested. We are not interested in the very largest $x$, e.g. the hardest 0.999999 quantile of our instances, because our problem of finding the ground state energy is NP hard, so the hardest instances are commonly assumed to be out of reach for current algorithms. In the next subsection we will see that the null-hypothesis cannot be rejected for our range of data and that the obtained GP distribution functions show excellent model fits, which justifies the use of GP distribution functions to fit our data.\\

\subsection{Fitting Procedure}
Here we introduce tests which could reject the null-hypothesis that our unknown distribution function satisfies the extreme value condition and show how to choose a high enough threshold $u$.

\begin{figure}[t]
  \centering
  \includegraphics[width=\linewidth]{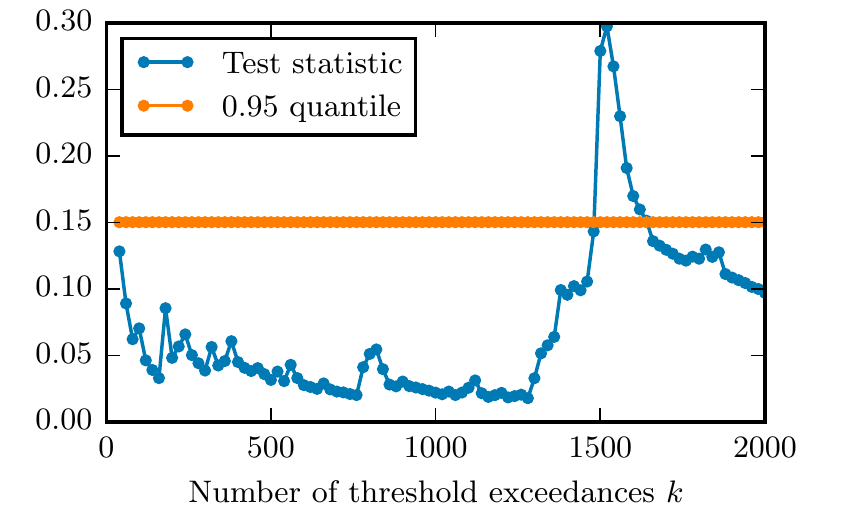}
  \caption{Dietrich, de Haan and H{\"u}sler's method to check the null-hypothesis 
  \cite{Dietrich2002} with $\eta=2$. Reject the null-hypothesis if the test statistic is larger than the 
  0.95 quantile of the limiting random variable. The full data set has 20'000 different instances. The 
  test statistic is calculated for different threshold values plotted according to the number of threshold
  exceedances $k$. The null-hypothesis has to be rejected for thresholds with more than
  1480 exceedances. This means we have to choose a threshold high enough such that there are less 
  than 1480 exceedances in this data set. Plotted is SA with $N=200$ spins and optimal number of sweeps $t_a^{\mathrm{opt}}$.} 
  \label{fig:evc_test_statistics}
\end{figure}

A common statistical approach to check the null-hypothesis is to estimate
the parameters of the GP distribution function from the exceedances over a threshold of 
the data. Based on these estimators, tail probabilities and quantiles are deduced. 
Then graphical methods are used to analyze the goodness-of-fit using for example
QQ and PP plots \cite[p.145]{Reiss2007}. These graphical methods are introduced in the next
subsection and the results didn't reject the null-hypothesis. 
In addition we used a method presented in \cite{Dietrich2002} to test the null-hypothesis,
that $F$ satisfies the extreme value condition with some additional (second order) condition.
A review is given in \cite[p.144-151]{Reiss2007} and in the paper by H{\"u}sler and Li 
\cite{Husler2006}, where they also provide a R program code. This code
is used in our paper to calculate the test statistics and the 0.95 quantiles of the 
limiting random variables by moment estimation for different number of 
threshold exceedances, an example is shown in Fig.~\ref{fig:evc_test_statistics}. 
The test statistic depends on the chosen threshold value. We saw that if the threshold value was 
too low, this test rejects the null-hypothesis, which provides us a hint on the smallest
possible threshold value. For all the chosen thresholds in this paper this test method doesn't 
reject the null-hypothesis.

\begin{figure}[t]
 \centering
  \includegraphics[width=\linewidth]{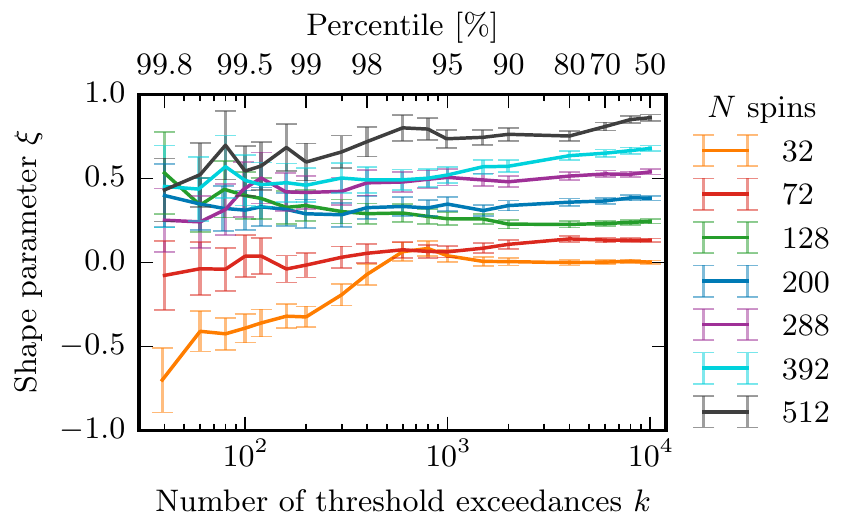}
 \caption{Fitted shape parameters $\xi$ with standard errors for problems with different number of spins $N$ depending on the threshold $u$. 
 This plot is used to determine a high enough threshold. $\xi$ should be constant 
 for thresholds higher than a high enough threshold. The threshold on the x-axis is given in percentiles or number of threshold exceedances $k$ from the total of 20000 random instances. 
 Drees \cite[p.237]{Drees2012} 
 mentions that it is a good idea to plot the estimated shape parameter
 as a function of $\log(k)$. Plotted is the data for SA with the optimal number of sweeps $t_a^{\mathrm{opt}}(N)$.
}
 \label{fig:figure_gpd_depending_on_u}
 \end{figure}

There are automatic ways of choosing a high enough threshold \cite[p.137]{Reiss2007}, however, 
we used two manual method which might achieve better fitting results:
First a range of thresholds is chosen and the exceedances over these thresholds 
are used to estimate the parameters of the GP distribution functions by the Maximum Likelihood Estimation (MLE) method 
\cite{Coles2001,Castillo2004}. We used MLE implementations for extreme value theory from the two
libraries ismev and evir \cite{ismev2012,evir2012} and 20'000 random problems are used for every algorithm, schedule and system size.
The estimated shape parameter $\xi$ as a function of the threshold is plotted in Fig.~\ref{fig:figure_gpd_depending_on_u}.
Because of the peak-over-threshold stability of
the GP distribution function, the estimated shape parameter should be constant for 
thresholds $\mu>u$ if the threshold $u$ is already high enough. Therefore, there are 
usually three regions in Fig.~\ref{fig:figure_gpd_depending_on_u}:
For a very high threshold there are only a few exceedances and 
hence the shape parameter fluctuates strongly. For a high threshold 
the shape parameter is stable around the true value. For a too low threshold 
there are too many exceedances and the extreme value condition might not
be satisfied, this can be detected if the shape parameter is not constant or
by testing the null-hypothesis, see Fig.~\ref{fig:evc_test_statistics}.
Now we have a range of threshold which seem to be high enough. To choose the exact threshold 
we used the PP and QQ plots to check which threshold shows the best model fit.

\subsection*{Evaluation of Model Fit}

\begin{figure}[t]
 \centering
 \includegraphics[width=\linewidth]{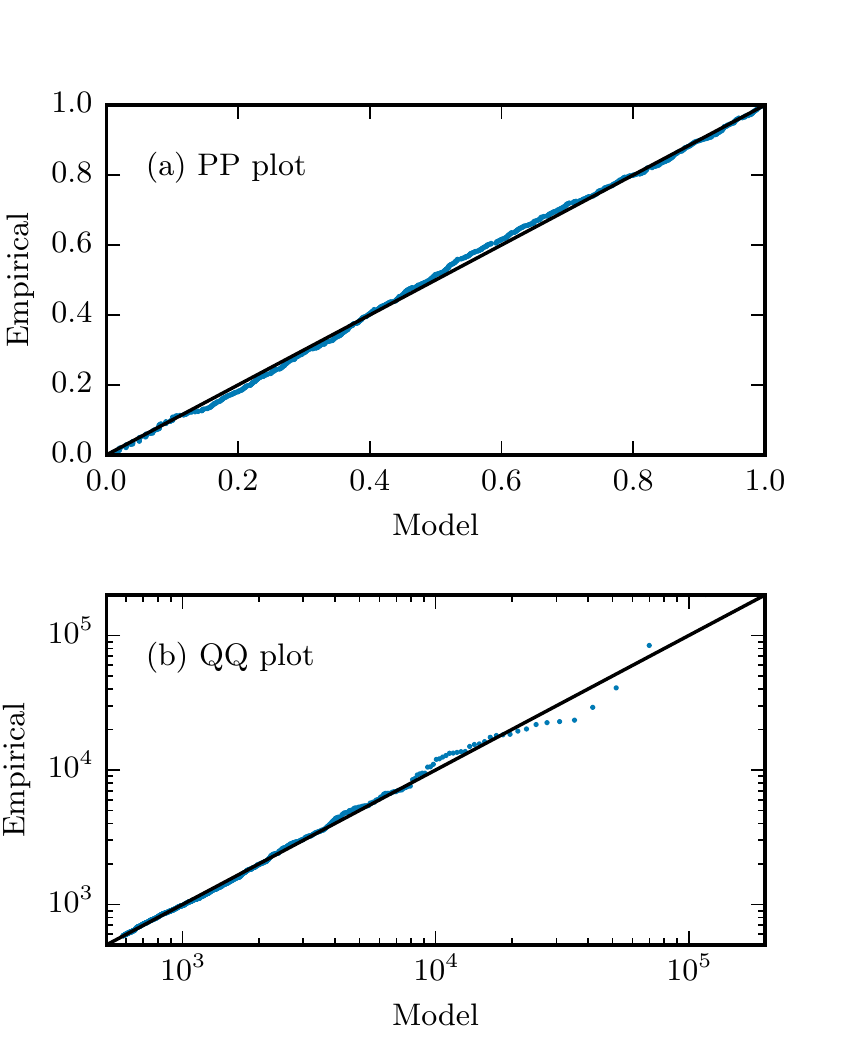}
  \caption{The exceedance distribution of $\tau$ over a high threshold for SA with 512 spins and optimal 
  number of sweeps $t_a^{\mathrm{opt}}$ is fitted to a GP distribution function. The goodness of the model fit is shown here with (a) PP and (b) QQ plot. As  the points are close to the unit diagonal it indicates that the estimated GP distribution function is an excellent fit for the data.
  }
 \label{fig:figure_qq and_pp_plot_sa}
\end{figure}

The goodness of the model fit can be graphically checked with PP and QQ plots.
Given an ordered sample of independent observations
\begin{equation}
\nonumber
x_{(1)}\le x_{(2)} \le \dots \le x_{(n)}
\end{equation}
the empirical distribution function $\tilde{F}$ is defined by
\begin{equation}
\nonumber
\tilde{F}(x)=\frac{i}{n+1}\quad \mathrm{for}\; x_{(i)}\le x \le x_{(i+1)}.
\end{equation}
Suppose $\hat{F}$ is an estimated distribution function of $F$, then a 
probability plot, also called PP plot, consists of the points
\begin{equation}
\label{eq:pp_plot}
\left\lbrace\left(\hat{F}\left(x_{(i)}\right),\frac{i}{n+1}\right): i=1,\dots,n\right\rbrace
\end{equation}
and a quantile plot, also called QQ plot, consists of the points
\begin{equation}
\label{eq:qq_plot}
\left\lbrace\left(\hat{F}^{-1}\left(\frac{i}{n+1}\right),x_{(i)}\right): i=1,\dots,n\right\rbrace,
\end{equation}
where $\hat{F}^{-1}$ is the estimated quantile function \cite[p.36]{Coles2001}.
If $\hat{F}$ is a good estimate of the distribution function $F$, then the points 
in a QQ plot and in a PP plot should lie close to the unit diagonal. An example is 
shown in Fig.~\ref{fig:figure_qq and_pp_plot_sa}. QQ plots turned out to much 
more sensitive about the threshold choice than PP plots. All PP and QQ plots 
show that the estimated model agrees excellently with the data.

\section{Running mean}

\begin{figure}[t]
 \centering
 \includegraphics[width=\linewidth]{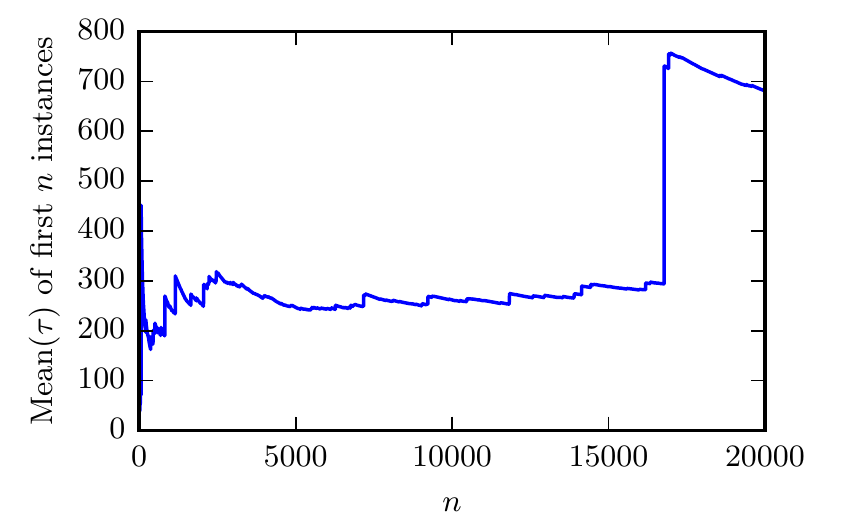}
  \caption{Plotted is the mean of $\tau$ of the first $n$ instances for SQA with $\beta=10$ and optimal annealing time $t_a^{\mathrm{opt}}$ for problems with 288 spins. We see that the mean is not converging using 20000 random instances and single instances dominate the mean, e.g. instance 16773 has $\tau=7333790$ and therefore moves the mean significantly upwards. This behavior can be expected as the estimated shape parameter of the GP distribution function is $\xi=1.10\pm0.07$ which means the mean of the GP distribution function is diverging.
  }
 \label{fig:running_mean}
\end{figure}

In the main paper we show that the tail of the distribution of $\tau$ for SQA is always heavy ($\xi\geq0$) for problems with $N\geq72$ spins. Sometimes the estimated shape parameter $\xi$ is even $\geq1$ indicating that the mean of the GP distribution function is infinite but, as there exists only a finite number of instances, the mean of the distribution function of $\tau$ is not divergent but dominated by the hardest instances. This can easily be seen by considering the running mean of $\tau$, i.e. the mean of $\tau$ for the first $n$ instances. See Fig.~\ref{fig:running_mean} which shows that the mean is not converging using 20000 random instances with 288 spins for SQA with $\beta=10$ and optimal annealing time $t_a^{\mathrm{opt}}$. This is expected because the estimated shape parameter $\xi$ is $1.10\pm0.07$.

\FloatBarrier
\section{DATA}
Here we show the results of the tail analysis and repeat the necessary details from the main text in order to be self-contained in the presentation of the data.

For every algorithm, schedule and system size $N$ we used 20000 random instances to estimate for each instance the single-run success probability $s$ to find the ground state by averaging over many repetitions. Then the mean number of repetitions $\tau=1/s$ required to find the ground state is calculated. The distribution function $F(\tau)$ is analyzed using extreme value theory: The exceedances over a high threshold $u$ are fitted to a GP distribution function, such that $F^{[u]}(\tau)\approx W_{\xi,u,\sigma_u}(\tau)$. In the tables below are the results for SA, SQA and MFA: $N$ is the number of spins, $u$ is the threshold, $k$ are the number of threshold exceedances, i.e. the number of instances that have $\tau>u$ from all the 20000 instances, $\xi$ is the estimated shape parameter and $\sigma_u$ is the estimated scale parameter of the GP distribution function $W_{\xi,u,\sigma_u}(\tau)$. Both $\xi$ and $\sigma_u$ are shown with the standard error obtained from the maximum likelihood estimation method.

\begin{table}[H] %EVT results SA optimal number of sweeps
\ra{1.3}
\begin{center}
\begin{tabular}{@{}lllll@{}}\toprule
$N$ \phantom{abc}& $k$\phantom{abcd} & $u$\phantom{abcd} & $\xi$\phantom{abcdefghijklm} & $\sigma_u$\\ \midrule
32   &200    &12.9   &$-0.32\pm 0.06$   &$3.61\pm 0.33$   \\
72   &1000  &21.3   &$0.06\pm 0.03$    &$7.79\pm 0.36$   \\
128 &1000  &37.0   &$0.26\pm 0.04$    &$14.6\pm 0.7$     \\
200 &1000  &58.4   &$0.35\pm 0.04$    &$29.6\pm 1.5$     \\
288 &998    &165    &$0.51\pm 0.05$    &$102\pm 6$         \\
392 &998    &317    &$0.52\pm 0.05$    &$260\pm 14$       \\
512 &400    &1208  &$0.72\pm 0.09$    &$1086\pm 103$   \\
\bottomrule
\end{tabular}
\label{tab:evt_results_sa_opt}
\caption{Parameters for SA with optimal annealing time $t_a^{\mathrm{opt}}(N)$. Note by comparing to Tab.~\ref{tab:evt_results_sa_s10000} that the estimated shape parameter $\xi$ is less
dependent on the number of sweeps than in the case of SQA. }
\end{center}
\end{table}

\begin{table}[H] %EVT results SA 10k sweeps for each system size
\ra{1.3}
\begin{center}
\begin{tabular}{@{}lllll@{}}\toprule
$N$ \phantom{abc}& $k$\phantom{abcd} & $u$\phantom{abcd} & $\xi$\phantom{abcdefghijklm} & $\sigma_u$\\ \midrule
32   &194    &1.5     &$-0.06\pm 0.06$ &$0.199\pm 0.019$  \\
72   &599    &2.4     &$0.31\pm 0.05$  &$0.564\pm 0.037$   \\
128 &400    &5.8     &$0.51\pm 0.08$  &$2.29\pm 0.20$       \\
200 &1000  &8.6     &$0.48\pm 0.05$  &$5.24\pm 0.29$       \\
288 &999    &20.1   &$0.59\pm 0.05$  &$14.5\pm 0.8$          \\
392 &599    &79.0   &$0.53\pm 0.07$  &$74.0\pm 5.5$          \\
512 &399    &373    &$0.86\pm 0.10$  &$324\pm 32$            \\
\bottomrule
\end{tabular}
\caption{Parameters for SA with $t_a=10000$ sweeps.}
\label{tab:evt_results_sa_s10000}
\end{center}
\end{table}

\begin{table}[H]%EVT results SQA beta=10, 10k sweeps
\ra{1.3}
\begin{center}
\begin{tabular}{@{}lllll@{}}\toprule
$N$ \phantom{abc}& $k$\phantom{abcd} & $u$\phantom{abcde} & $\xi$\phantom{abcdefghijkl} & $\sigma_u$\\ \midrule
32   &200   &1.6      &	$0.63\pm 0.12$ &$0.813\pm 0.110$ \\
72   &200   &26.0    &$1.27\pm 0.16$ &$23.0\pm 3.4$        \\
128 &400   &113     &$1.71\pm 0.14$ &$238\pm 28$          \\
200 &400   &1381   &$1.85\pm 0.14$ &$2931\pm 354$      \\
288 &318   &36281 &$1.86\pm 0.15$ &$88127\pm 1485$  \\
\bottomrule
\end{tabular}
\label{tab:evt_results_sqa_b10_s10000}
\caption{Parameters for SQA with $\beta=10$ and $t_a=10000$ sweeps. Note that the ground state of the hardest instance with $N=288$ spins was never found in 2'373'660'226 repetitions. We therefore carried out the analysis using an upper bound on the success probability of $s=4.21\times10^{-10}$ for this instance.}
\end{center}
\end{table}

\begin{table}[H]%EVT results SQA beta=10, opt. sweeps
\ra{1.3}
\begin{center}
\begin{tabular}{@{}lllll@{}}\toprule
$N$ \phantom{abc}& $k$\phantom{abcd} & $u$\phantom{abcd} & $\xi$\phantom{abcdefghijkl} & $\sigma_u$\\ \midrule
32   &400   &7.9     &$0.07\pm 0.06$ &$2.10\pm 0.16$  \\
72   &200   &23.2   &$0.31\pm 0.09$ &$8.00\pm 0.89$  \\
128 &400   &47.3   &$0.47\pm 0.07$ &$26.1\pm 2.3$    \\
200 &400   &219    &$0.76\pm 0.09$ &$170\pm 16$      \\
288 &1000 &603    &$1.10\pm 0.07$ &$577\pm 37$      \\
\bottomrule
\end{tabular}
\label{tab:evt_results_sqa_b10_opt}
\caption{Parameters for SQA with $\beta=10$ and optimal annealing time $t_a^{\mathrm{opt}}(N)$.}
\end{center}
\end{table}

\begin{table}[H]%EVT results SQA beta=4, opt. sweeps
\ra{1.3}
\begin{center}
\begin{tabular}{@{}lllll@{}}\toprule
$N$ \phantom{abc}& $k$\phantom{abcd} & $u$\phantom{abcd} & $\xi$\phantom{abcdefghijklm} & $\sigma_u$\\ \midrule
32   &400   &6.5      &	$-0.01\pm 0.05$ &$1.71\pm 0.13$ \\
72   &400   &15.7    &$0.19\pm 0.06$   &$5.62\pm 0.41$ \\
128 &398   &48.3    &$0.40\pm 0.07$   &$22.1\pm 1.8$   \\
200 &200   &318     &$0.54\pm 0.11$   &$187\pm 23$     \\
288 &400   &648     &$0.75\pm 0.09$   &$629\pm 59$     \\
\bottomrule
\end{tabular}
\label{tab:evt_results_sqa_b4_opt}
\caption{Parameters for SQA with $\beta=4$ and optimal annealing time $t_a^{\mathrm{opt}}(N)$.}
\end{center}
\end{table}

\begin{table}[H]%EVT results MF beta=4, opt sweeps
\ra{1.3}
\begin{center}
\begin{tabular}{@{}lllll@{}}\toprule
$N$ \phantom{abc}& $k$\phantom{abcd} & $u$\phantom{abcd} & $\xi$\phantom{abcdefghijklm} & $\sigma_u$\\ \midrule
32   &400   &7.6      &$	-0.03\pm 0.05$  &$1.96\pm 0.14$    \\
72   &198   &42.4    &$0.17\pm 0.08$     &$10.5\pm 1.1$       \\
128 &398   &140     &$0.36\pm 0.07$     &$58.9\pm 4.8$       \\
200 &400   &226     &$0.60\pm 0.08$     &$139\pm 12$         \\
288 &400   &998     &$0.90\pm 0.10$     &$812\pm 80$         \\
392 &198   &9302   &$0.96\pm 0.14$     &$10071\pm 1779$ \\
\bottomrule
\end{tabular}
\label{tab:evt_results_mf_b4_opt}
\caption{Parameters for MFA with optimal $\beta=4$ and optimal annealing time $t_a^{\mathrm{opt}}(N)$.}
\end{center}
\end{table}

\begin{table}[H] %EVT results MF beta =10, opt sweeps
\ra{1.3}
\begin{center}
\begin{tabular}{@{}lllll@{}}\toprule
$N$ \phantom{abc}& $k$\phantom{abcd} & $u$\phantom{abcd} & $\xi$\phantom{abcdefghijklm} & $\sigma_u$\\ \midrule
32   &298   &23.5    &$	-0.19\pm 0.05$  &$6.92\pm 0.53$  \\
72   &398   &54.5    &$0.06\pm 0.06$     &$18.6\pm 1.4$       \\
128 &100   &256     &$0.58\pm 0.15$     &$105\pm 18$       \\
200 &398   &459     &$0.86\pm 0.09$     &$428\pm 42$       \\
288 &399   &3419   &$1.35\pm 0.12$     &$4312\pm 477$   \\
\bottomrule
\end{tabular}
\label{tab:evt_results_mf_b10_opt}
\caption{Parameters for MFA with $\beta=10$ and optimal annealing time $t_a^{\mathrm{opt}}(N)$.}
\end{center}
\end{table}

\end{document}